\documentclass[preprint]{aastex}
\usepackage{amsmath,amssymb,amstext}

\usepackage{tablefootnote}
\usepackage[breaklinks,colorlinks,citecolor=blue,linkcolor=magenta]{hyperref} 

\usepackage[all]{hypcap} 

\usepackage{aas_macros}
\usepackage{natbib}
\shorttitle{Modeling HL Tau}
\shortauthors{Jin et al.}

\begin{document}

\title{Modeling Dust Emission of HL Tau Disk Based on Planet-Disk Interactions}
\author{Sheng Jin \altaffilmark{1,2}, Shengtai Li \altaffilmark{2}, Andrea Isella \altaffilmark{3},
Hui Li \altaffilmark{2}, Jianghui Ji \altaffilmark{1}}

\altaffiltext{1}{Key Laboratory of Planetary Sciences, Purple Mountain Observatory, 
Chinese Academy of Sciences, Nanjing 210008, China}

\altaffiltext{2}{Theoretical Division, Los Alamos National Laboratory, Los
Alamos, NM 87545}

\altaffiltext{3}{Rice University, Houston, TX}

\begin{abstract}
We use extensive global two-dimensional hydrodynamic disk gas+dust simulations 
with embedded planets, coupled with three dimensional radiative transfer calculations, 
to model the dust ring and gap structures in the HL Tau protoplanetary disk observed 
with the Atacama Large Millimeter/Submillimeter Array (ALMA). 
We include the self-gravity of disk gas and dust components and  
make reasonable choices of disk parameters, assuming an already settled dust
distribution and no planet migration.
We can obtain quite adequate fits 
to the observed dust emission using three planets with masses 0.35, 0.17, and 0.26 $M_{Jup}$
at 13.1, 33.0, and 68.6 AU, respectively. 
Implications for the planet formation as well as the limitations of 
this scenario are discussed.
\end{abstract}

\keywords{protoplanetary disks --- planet-disk interactions --- techniques: interferometric --- submillimeter: planetary systems}

\maketitle

\section{Introduction}

ALMA observations of HL Tau, a young ($\leq$ 1-2 Myr ) star in Taurus,  
revealed bright and dark rings in the millimeter-wave continuum emission 
\citep{Partnership2015}. One interesting question is whether these structures 
are created by the embedded planets, hence yielding potentially
important clues on planet formation in such disks. 

Other mechanisms besides the effects of planets have
been suggested, including zonal flows \citep{Ruge2013}, Rossby wave instability 
\citep{Lovelace1999, Li2000, Li2005, Regaly2012,Pinilla2012},  
rapid pebble growth around condensation fronts \citep{Zhang2015},
and dust dynamics \citep{Gonzalez2015}.
Some observational features suggest however that the pattern of 
rings shown in HL Tau disk may be produced by planets.
First, the spectral index derived from the ALMA images suggests that
the dark rings are optical thin whereas the bright regions are optically thick, 
therefore the dark rings are real gaps in the dust distribution \citep{Partnership2015}.
Second, the increase of the eccentricity of the rings at large orbital radii 
\citep{Partnership2015} is consistent with the fact that the orbital eccentricities of 
exoplanets increase with orbital radii \citep{Butler2006,Shen2008,Zhang2014}.
Third, the dust size constrained by the polarized emission is around 150 $\mu m$ 
\citep{Kataoka2015}, which means the structure of multiple rings should also 
exist in the gas disk because dust at this size should be well coupled with gas.
\citet{Pinte2015} conclude that the depletion of dust at each of the deepest 
gaps is up to 40 $M_{\oplus}$, which is close to the point of runaway gas accretion. 

Models of disks that couple the dynamics of gas, dust and planets are crucial to 
interpret the observed patterns in the HL Tau system \citep[e.g.,][]{Dong2014, Zhu2015, 
Dipierro2015}. Various models suggest that the disk mass of HL Tau is 
$\sim$ 0.03-0.14 $M_{\odot}$ 
\citep{Robitaille2007,Guilloteau2011,Kwon2011,Kwon2015}, 
and the estimated stellar mass of HL Tau is $0.55$ to $1.3~ M_{\odot}$ 
\citep{Sargent1991,Close1997,White2004,Partnership2015}.
The high disk mass might affect the disk stability and 
the resonance locations of potential planets \citep{Tamayo2015}. 
A recent study by \citet{Dipierro2015} 
used a three-dimensional gas+dust two-fluid SPH code that includes embedded planets.
Though they were able to reproduce the pattern of bright and dark rings, 
they adopted a disk mass of only $0.0002~ M_{\odot}$ 
within 120 AU and under predicted by about a factor 20 the observed 
millimeter flux density.

In this paper, we perform quantitative fitting to the
ALMA observations of HL Tau in terms of its millimeter flux density and spatial variations. 
We address the question whether the observed
features of HL Tau by ALMA can be produced by planet-disk interactions. 
We describe our approach and model set-up in Section \ref{sec:model}, 
summarize our main results in Section \ref{sec:results}, and 
discuss their implications in Section \ref{sec:conclu}.

\section{Models}
\label{sec:model}

The model adopted to analyze HL Tau observations makes use of the 
LA-COMPASS code \citep[][]{Li2005,Li2009,Fu2014} to calculate the planet-disk 
interaction, and of radiative transfer code {\em RADMC-3D}\footnote{http://www.ita.uni-heidelberg.de/~dullemond/software/radmc-3d/}
to calculate the disk temperature and emission at 1 mm. 
Our aim is not to obtain a perfect fit to the ALMA observation of HL Tau
but instead to investigate in a quantitative way whether the planet-disk interactions
might explain the observed features. 


\subsection{Disk gas+dust+planet model}


We initialize our hydrodynamic simulation using the results from \cite{Kwon2011}. 
The disk extends from 2.4 AU to 156 AU, and is simulated using 
a polar grid containing 3072 and 768 cells 
in the radial and azimuthal direction, respectively. 
We adopt a disk surface density described by  
\begin{equation}
\label{eq:sigmag}
\Sigma_g(r) = \Sigma_0 \Big(\frac{r}{r_0}\Big)^{0.23}\exp[-(r/r_c)^{2.23}]~~,
\end{equation}
where 
$r_c = 79$ AU is a characteristic radius \citep[e.g.,][]{Andrews2009} and
$r_0  = 10$ AU.
The mass of the star is $M_*= 0.55 M_{\odot}$. 
Although \cite{Kwon2011} finds a disk mass of $0.135 M_{\odot}$, 
we find that the disk with this mass quickly 
becomes gravitationally unstable under small perturbations. 
Hence we reduce the disk
mass to be $M_{\rm{disk}} \approx 7.35 \times 10^{-2}$
$M_{\odot}$, which leads to $\Sigma_0 = 4.83\times
10^{-4} [M_*/(10 {\rm AU})^2] \approx 23.61$ g/cm$^2$. 
The whole disk is assumed to have a constant Shakura-Sunyaev 
viscosity parameter $\alpha = 10^{-3}$. 
The locally isothermal sound speed is chosen as: 
\begin{equation}
\Big(\frac{c_s}{v_\phi} \Big)(r) = 0.05 \Big(\frac{r}{r_0}\Big)^{0.285}~~.
\end{equation}     

We assume that the initial dust surface density
$\Sigma_d(r)$ follows $\Sigma_g(r)$, with 
an initial dust-to-gas ratio of $1\%$. We have performed gas+dust 
two-fluid simulations using different dust particle sizes 
ranging from $\mu$m to millimeter, though only one dust size is used in a single hydro 
simulation. Such gas+dust distributions are used to calculate the  dust temperature
as well as the dust emissions which are ultimately compared with ALMA observations. 
For comparison with the ALMA images, we have used the 
$0.15$ millimeter size dust particles from
gas+dust two-fluid simulations (see below for dust models). 
The initial Stokes number of $0.15$
millimeter  size dust particles is $< 0.01$ within $\sim 100$ AU. 
For such small Stokes number, 
we adopt short friction time approximation \citep{Johansen2005} to circumvent 
problems with small time steps. 


The disk self-gravity and indirect acceleration term from the central
star are always included. 
An extended disk that includes inner disk $r\in
[0.01,0.24]$ and outer disk $r\in [15.6,156]$ is used in disk self-gravity 
calculations. 
We emphasize that the effects
of both disk self-gravity and indirect acceleration term are important due to
the relatively massive disk in HL Tau. In fact, simulations without including such
terms could give erroneous results. 
A fixed boundary condition is used for gas at both inner and
outer disk boundaries.  For
dust, we adopt an inflow (outflow) boundary
condition at the inner (outer) boundary.

\citet{Partnership2015} identify seven pairs of dark and bright rings in the $1.0$ mm 
image of the HL Tau disk, and among them four prominent dark rings could be considered 
as gaps, i.e., the dark rings D1, D2, and the adjoining D5 and D6. 
Based on a large body of previous literature on gap 
opening and formation as functions of disk gas temperature, planet mass, disk viscosity
\citep[e.g.,][]{Crida2006,Li2009},
we have performed tens of disk+planet hydro simulations in order to 
determine the likely planet mass values that could roughly match the multiple 
gap widths and depths 
as shown in the ALMA observations. Based on these simulations, we adopt a nominal model 
({run0}) which
has a set of planet mass parameters as $M_p \approx 0.35, 0.17,$ and $0.26 M_{\rm Jup}$ 
at $13.1, 33.0$, and $68.6$ AU, respectively. 
These three planets are fixed at their orbital locations, 
and planetary radial migration is turned off. 

To explore how our results could vary with some key parameters, we have carried out
many additional runs. In Table \ref{tab:simu}, we list a few simulations where we reduce 
the masses for all three planets by half ({run1}), disable the disk self-gravity ({run2}),
and reduce the disk viscosity ({run3}).  Other changes such as the disk 
surface density and mass are not presented here. 
Most of the results presented in this paper
use the snapshot at  4000 orbits ($\sim 170,556$ years at 10 AU) to 
generate simulated observations, except for {run1} where the snapshot at 
3400 orbits is used.

\begin{table}
\caption{Simulation Parameters}
\label{tab:simulist}
\begin{tabular}{lccccc}
\hline
\hline
Model & DSG    	  & $M_{\rm p} (M_{\rm Jup})$    & $\alpha$ & orbit no. \\
\hline
run0  & Y  & 0.35, 0.17, 0.26    & 1e-3 & 4000  \\
run1   & Y  & 0.17, 0.09, 0.13    &  1e-3 & 3400  \\
run2    & N  & 0.35, 0.17, 0.26    & 1e-3 & 4000 \\
run3   & Y  & 0.35, 0.17, 0.26    &  2e-4 & 4000 \\
\hline
\vspace{0.05cm}
\end{tabular}
\label{tab:simu}
\end{table}

\subsection{Dust model, radiative transfer model and image generation}

The dust opacities adopted in this work were calculated as in \cite{Isella2009} 
by assuming that dust grains are compact spheres made of astronomical silicates  \citep{wd01},
organic carbonates \citep{z96},
and water ice, with fractional abundances as in \cite{p94}.
Single grain opacities were averaged
on a grain size distribution to obtain the mean opacity used in the
radiative transfer model. We  did that by adopting a typical MNR power-law size
distribution \citep{mnr77}, $n(a) \propto a^{-3.5}$, between a minimum grain size of
$5 \times 10^{-4}$ mm and a maximum grain size of 10 mm.
The resulting dust opacity at the wavelength of 1~mm is 4.6 cm$^2$ g$^{-1}$,
and it is dominated by dust grains with sizes between 0.1 and 0.2 mm. Based on
this assumption, in our hydrodynamic models we adopt dust grains
with a size of 0.15 mm as best tracers of the dust emission at 1 mm.

To calculate synthetic maps of the disk continuum emission, we first need 
to calculate the dust temperature throughout the disk. This is controlled by 
micron-size dust particles that are well coupled with gas.
We adopt a dust-to-gas ratio of $0.01$ and use the gas surface density from
simulations to interpolate the micron-size dust density distribution on a 3D 
spherical grid along with a scale height profile of,
$h_{\rm \mu m-dust}(r) = 1.0 \textrm{AU} \times (r/20 \textrm{AU})^{1.25}$.
We set up a thermal Monte Carlo run to calculate the dust temperature using the 
radiative transfer code {\em RADMC-3D}.
The calculated dust temperature profile shows the typical two-layer vertical structure 
of passive irradiated circumstellar disks \citep{Dullemond2002}.
In the surface layer, the temperature decreases from $\sim$ 280 K at 5 AU to 
$\sim$ 75 K at 100 AU. In the midplane, the temperature decreases 
from $\sim$ 110 K at 5 AU to $\sim$ 20 K at 100 AU.
Next, using the obtained dust temperature, we calculate the millimeter 
thermal emission based on the density and opacity of our dust disk. 
We convert the surface density of 0.15 millimeter size dust from two-fluid simulations 
to a 3D spherical density profile using a  scale height of,
$h_{\rm mm-dust}(r) = f_{sett} \times h_{\rm \mu m-dust}(r)$,
where $f_{sett} = 0.1$ is a parameter that accounts for the settling of 0.15 millimeter size dust
towards the mid-plane.
This leads to a scale height of $\sim$ 0.75 AU at 100 AU for relatively large dust particles, 
which is consistent with the findings by \cite{Pinte2015}.
Using our gaseous disk parameters, 
the estimated settling time of $0.15$ mm dust is within 10$^5$ years 
at the scale height of gas disk,
thus it is a reasonable choice of dust scale height,
considering the age of the HL Tau disk.
We then generate the synthetic images at the millimeter wavelength using the ray tracing 
method of {\em RADMC-3D}.
As a last step, we  ``observe''  our models and compare them  
with the ALMA observations in the Fourier domain, following the 
procedure described in \cite[][]{Isella2009}.

\begin{figure*}
\includegraphics[width=17.0cm]{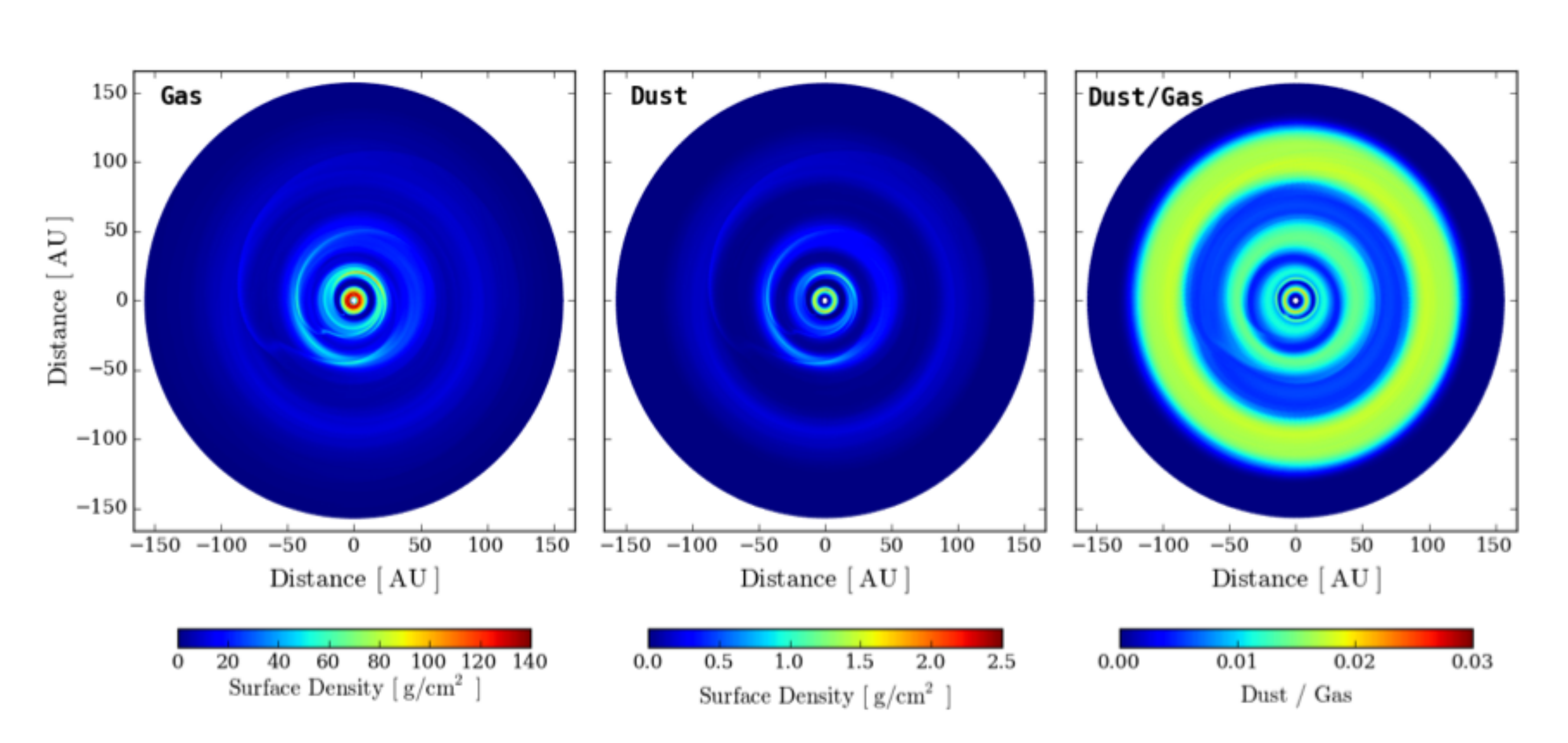}\\
\caption{Surface density of the gas (left) and of  0.15~mm size dust grains (center) 
after 4000 orbits for our nominal model that has three planets with masses of 0.35, 0.17, and 
0.26 $M_{\rm Jup}$ at 13.1, 33.0, and 68.6 AU, respectively. 
The right panel shows the dust-to-gas ratio.} 
\label{fig:hydro}
\end{figure*}

\begin{figure*}
\includegraphics[width=17.0cm]{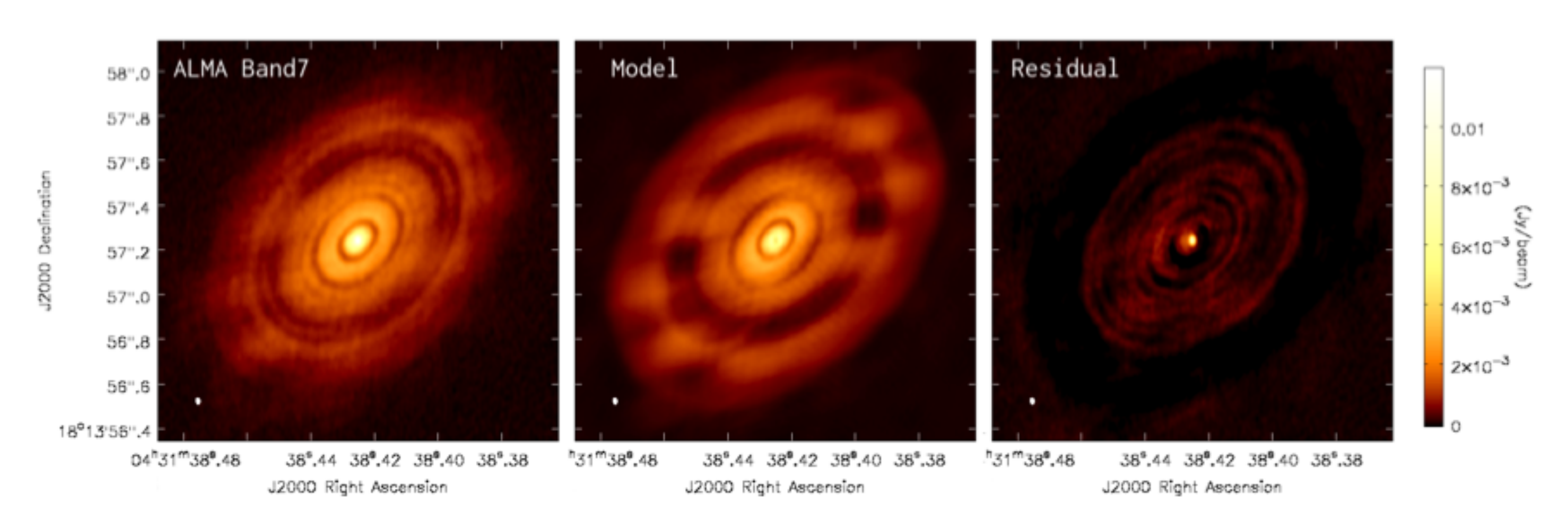}
\caption{
Comparison between the ALMA map of the HL Tau disk  ({\em left}) and 
the simulated observation of our nominal model ({\em middle}). The right panel shows 
the residual, i.e. the difference between the observation and our model.
The white point in the lower left corner shows the synthesized beam.}
\label{fig:simobs}
\end{figure*}



\section{Results}
\label{sec:results}

\subsection{The nominal model}

Figure \ref{fig:hydro} shows the surface density distributions of gas and 0.15
millimeter size dust in 
our nominal model. The gas and dust disks exhibit similar morphology,
showing three gaps and three higher density rings/bands of gas
created by the disk-planets interactions, along with visible spiral arm features.
The rings/bands in the dust distribution show asymmetries associated with the
spiral density features produced by the planets, especially in the inner and middle rings.
The dust to gas surface density ratio (initially at $0.01$) 
shows the evolution of dust+gas distributions
while dust is subject to gas drag and is concentrated at high pressure region, 
causing significantly enhancements at the locations of rings. Dust filtration effects 
are expected in the gaps because the larger size dust particles are more concentrated
in the rings. 
The deficit of dust at large radii $> 120$ AU is due to the fact that dust particles
drift radially inward and there is no supply of dust particles through the outer 
boundary.

Since HL Tau system is relatively young, the viscous 
timescale at disk radii $> 10$ AU is longer than the age of the disk. So, the disk
is not expected to be in quasi-steady state. 
When monitoring the evolution of the ring and gap structures over time, we indeed find
that the accretion rate in simulations is not steady and the depths of the gaps 
are still evolving (say between 4000 and 7000 orbits), though
understandably the first gap is approaching a steady state faster.

Figure \ref{fig:simobs} shows the ALMA observation of HL Tau, 
the synthetic observations of our nominal model, and a residual map 
obtained by subtracting the model from the observations in the Fourier domain.   
Overall, our model reproduces the observations quite well. 
(Good agreement is obtained between our model and ALMA Band 6 and 7 observations,
and we are only showing the Band 7 image due to its higher spatial resolution.)
What is most remarkable is that the flux density level between the model and observations 
can be matched so closely, given the fact that we have used a generic smooth gaseous 
disk with a uniform initial dust to gas ratio throughout such a disk. 
The simulated observation of our model can explain pretty well the three major gaps 
at $\sim$ 13, 33, and 68 AU, but it does not reproduce the detailed structures of multiple 
rings at larger radii due to our choice of using only one planet at 68 AU.
The large difference at the center region shown in the residual is due to our choice of
$2.4$ AU as the inner disk boundary.
Interestingly, the spiral arms and the asymmetry in the rings that are visible in 
the gas and dust distributions (cf Figure \ref{fig:hydro}) are mostly smeared out in the 
simulated observation due to the high optical depth of the continuum emission
at 850 $\mu$m. 

One of the key findings from the ALMA observations is that rings and gaps are
somewhat eccentric. To quantify this feature in our model image, by using a face-on configuration,
we find the local maxima (for the rings) or minima (for the gaps) in the azimuthal 
direction and fit the locations of these points with ellipses.  We obtain the 
eccentricities of gaps as 0.246, 0.274 and 0.277, respectively, whereas the
eccentricities of rings as 0.191, 0.079 and 0.157, respectively. This reproduces the
trend seen in the actual ALMA observations, though quantitative 
comparisons are somewhat difficult. 
This reproduces the trend seen in the actual ALMA observations.
Because we have calculated the eccentricities using our
simulations which have much higher 
spatial resolutions than ALMA observations, simulations can yield
more pronounced non-axisymmetric features on smaller scales, 
this makes more quantitative comparisons with observations difficult.

\begin{figure*}
\includegraphics[width=15cm, clip=True]{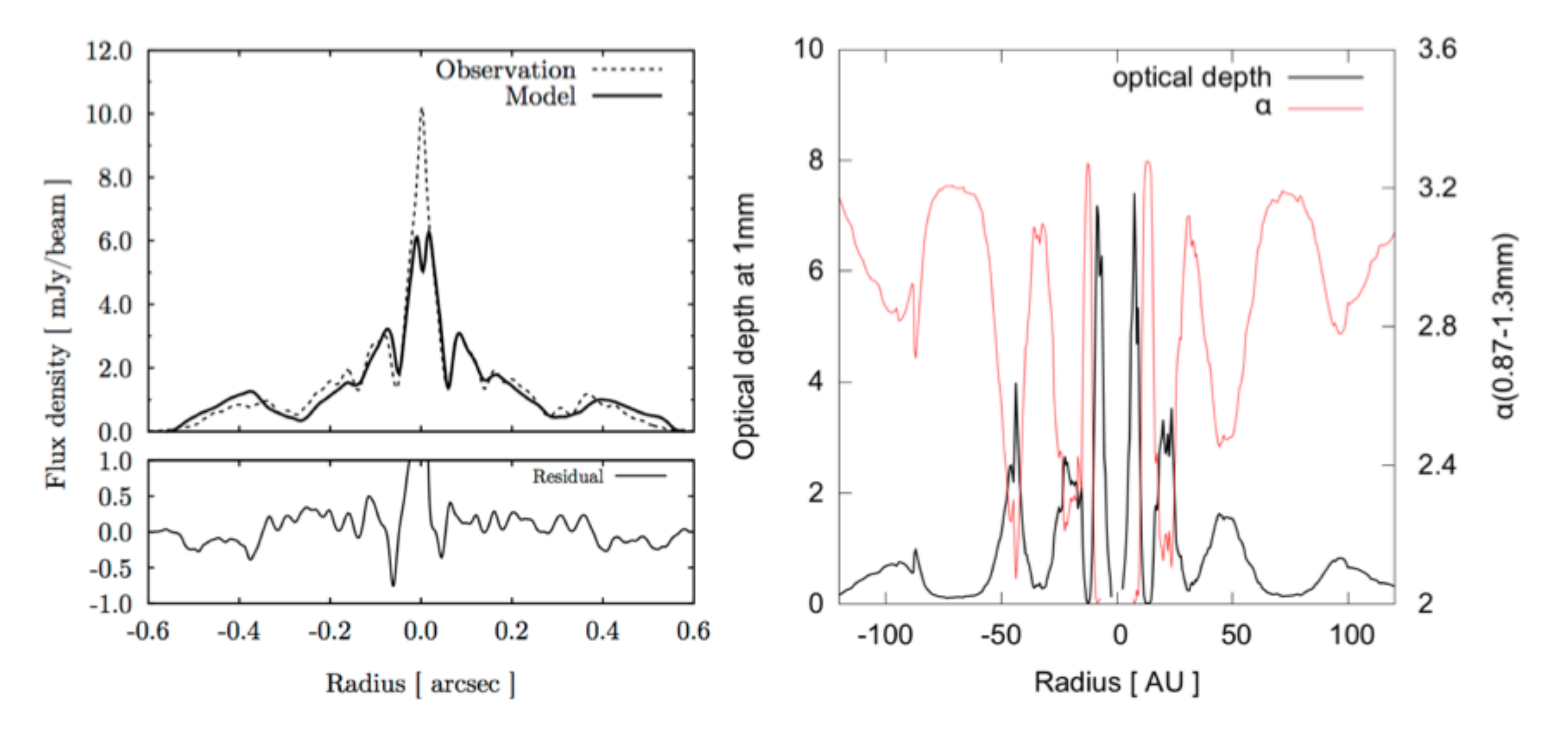}
\caption{Left: Comparison of the flux densities of our model and the ALMA Band 7 observation
using a line cut along the major axis of the disk. The bottom panel shows the residual. Right: 
Optical depth at 1 millimeter (dark line) and millimeter spectral index $\alpha$ of the model emission  
using a line cut along the major axis of the disk. The dark ring regions are mostly optical thin 
whereas other parts of the disk are optically thick.}
\label{fig:nominal}
\end{figure*}

The left panel of Figure \ref{fig:nominal} shows the radial profile of the 
intensity along the major axis of the disk corresponding 
to a position angle of $138^{\circ}$.
The flux densities of our model match the observation well, especially in the 
region around the first and second ring+gap.
At the third ring and beyond, the relative difference 
between our model and the observation increases. Again, it is worthwhile emphasizing
that the relative heights among different regions were all obtained self-consistently through
two-fluid gas+dust simulations along with the radiative transfer calculations. 

The right panel of Figure \ref{fig:nominal} shows the optical depth $\tau$ and the 
spectral index $\alpha$ of the dust emission calculated between 0.87~mm and 1.3~mm along 
the disk major axis.  The dark rings are (partially) optically thin while the
rest of the disk is mostly optically thick. For $\tau \ll 1$,  $\alpha$ depends on 
the slope of the dust opacity $\beta$ ($\alpha \sim 2+\beta$). Vice versa,  for $\tau \gtrsim 1$, 
$\alpha$ approaches 2. The radial profile of $\alpha$ predicted by our model is 
also in remarkable agreement with the observations \citep[see Figure 3 of ][]{Partnership2015}.

Finally, note that since we used only one dust size in our simulations, 
the spectral index variation in our model is only determined by the 
properties of the dust disk (such as optical depth). In real observations, 
the radial variation of spectral index might also depend on 
the dust properties (such as size evolution) through the disk
 \citep{Guilloteau2011,Testi2014}.

%

\begin{figure*}
\includegraphics[width=17.0cm]{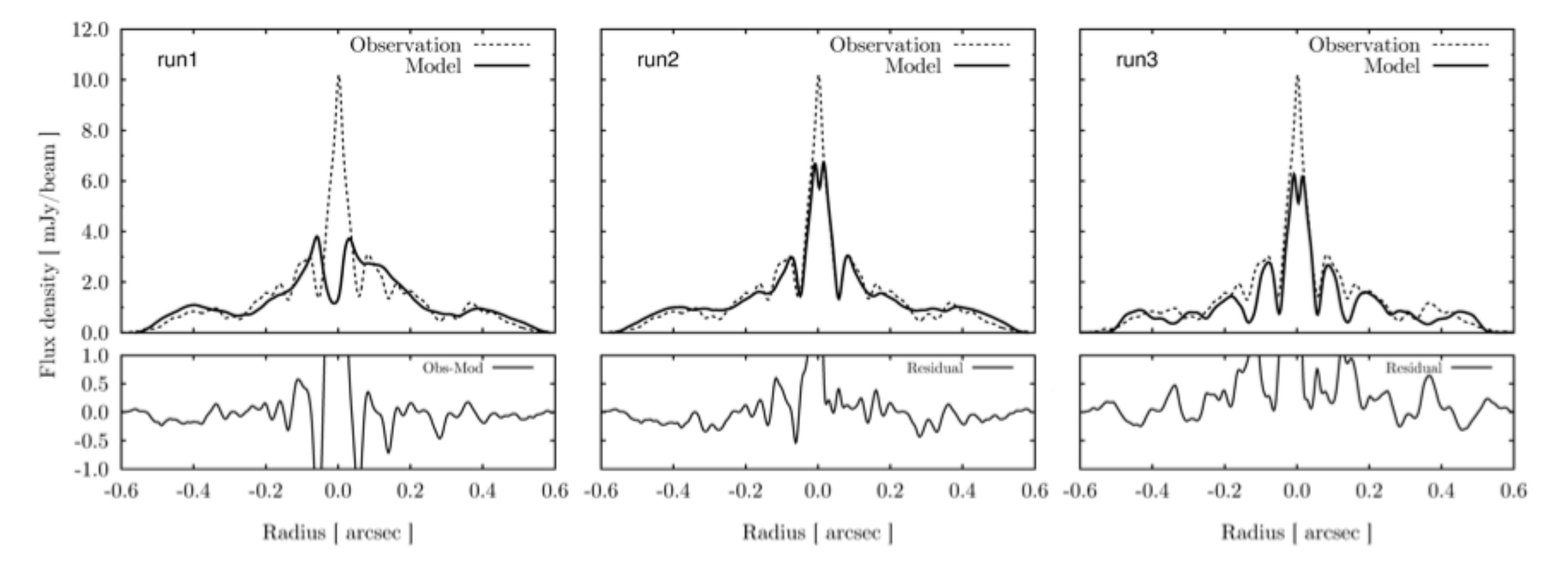}
\caption{
Comparison of the flux densities along the disk major axis for  
model run1, 2, and 3,  and the ALMA observation.  
} 
\label{fig:compmodel}
\end{figure*}

\subsection{Influence of parameters}

In Figure \ref{fig:compmodel} we compare the other three models 
with the ALMA Band 7 image.
Similar with Figure \ref{fig:nominal}, we draw cross-cuts of the simulated observations 
of run1, 2, and 3 along the major axis of the disk and compare the obtained 
flux density profiles with the observation.
Run1, where the planet masses are half of the nominal model, 
fails to produce the second gap and the depth of the first gap is too shallow compared
to the observation. This is expected since less massive planets can not open deep gaps.
Run2, where  the disk self-gravity is turned off, fails to produce the third gap. 
This is because, as the third planet gradually creates a gap, the gravitational force 
of the disk interior of the planet will try to pull the outer disk inward, further promoting 
the formation of a partial gap. Without such disk self-gravity, the gap is much shallower. 
In run3,  where the disk viscosity is reduced by a factor of 5, the gaps are too deep 
because the spiral shocks produced by planets in low viscosity disks are more effective 
in clearing the regions surrounding the planets. 

Note that the results shown in Figure  \ref{fig:compmodel} do not mean that we could not 
find a suitable fitting when parameters are different from our nominal model. On the contrary,
it is likely that a right combination of planet masses,
disk viscosity, temperature, mass, as well as the disk evolution time could be found
to produce a reasonable fit
to the current ALMA observation. The parameter studies in this subsection
are meant to help us to gauge how sensitive the final outcome is to various parameters.

\section{Summary and Discussion}
\label{sec:conclu}

We have performed 2D hydrodynamic simulations of disk gas+dust evolution
including disk self-gravity and embedded planets to model the HL Tau disk.
The simulation results are then coupled with 3D radiative transfer calculations 
to be compared with the observation of Partnership et al. (2015).
By comparing synthetic emission maps and observations in the Fourier domain, we 
have tested whether the observed multiple rings and gaps  could be
generated by the disk-planet interactions. Through a large number of simulations, we
find that the observed features can indeed be mostly reproduced by planet-disk interactions.
Without any fine-tuning, we find that disk model containing three planets with masses of
 $\sim 0.35, 0.17, 0.26 M_{\rm Jup}$, located at $13.1, 33.0$, and $68.6$ AU, respectively, 
is able to reproduce the the radial profile of the flux density, as well as, 
its spectral index. The rings and gaps from the model are also eccentric,
consistent with the ALMA observations. Furthermore, our fitted model parameters
favor a relatively massive disk and an initial dust to gas ratio about $0.01$.
The effects of self-gravity are essential in reproducing the profiles for gaps and rings.

There are, however, several outstanding issues that deserve further studies. 
First, the three planets in our model are assumed to stay on fixed radii. When 
allowed to migrate, they eventually move out of the current locations. Although
the observed gaps seem to be in resonance, it is difficult to understand why
planets will be in such resonances, given the fact that their radial migration speeds
tend to be fast in a high-mass disk environment \citep[cf.][]{Zhang2014}. 
Second, we find that the parameters we adopted for the disk and planets 
are barely able to keep the disk stable and with low eccentricity 
(we will present additional results in a future
publication). The disk self-gravity, as well as the indirect acceleration term,
are playing an important role in such disks. In general, contrary to the quasi-axisymmetry of the 
observed HL Tau image, the disk and planet orbits are expected to be more eccentric
for such disk and planet masses. 
Third, previous studies have suggested that HL Tau could be an FU Ori system
in quiescence \citep{L94}, which means that mass must be accumulating in the disk.
In addition, the presence of jet/outflows suggests that the angular momentum transport
processes in this disk are quite complicated. This non-steady situation, coupled with
uncertainties in the lifetime of planets, makes it difficult to constrain the planet mass 
more precisely. 

Nonetheless, the high-resolution ALMA observations of HL Tau system have enabled
detailed modeling of this system,
and we expect observations at longer wavelengths of this system 
from JVLA will further constrain the models. Using the same model in this paper 
but with $1.5$ mm dust grain size (the dust that dominates the centimeter emission), 
we find that the rings we saw at 1 mm become much narrower at the 1 cm wavelength
because the effect of dust concentration at high pressure regions is 
more prominent for larger grains.
We also find that these three bright rings are still optical thick at 1 cm wavelength.
Future observations of the morphology of HL Tau system at longer wavelengths 
can provide additional constraints for understanding dust dynamics and planet formation.

\acknowledgments{
This paper makes use of the ALMA data.
The Joint ALMA Observatory is operated by ESO, AUI/NRAO and NAOJ. 
Support by LANL's LDRD, UC-Fee and IGPPS programs are gratefully 
acknowledged. J.H. and S.J. appreciate the support by National Natural
Science Foundation of China (grants no. 11273068, 11473073, 11503092),
the Strategic Priority Research Program-The Emergence
of Cosmological Structures of the Chinese Academy of
Sciences (grant no. XDB09000000), the innovative and interdisciplinary
program by CAS (grant no. KJZD-EW-Z001), the Natural
Science Foundation of Jiangsu Province (grant no. BK20141509)
and the Foundation of Minor Planets of Purple Mountain Observatory.
A.I. acknowledge support from NSF (grant no. 1535809) and NASA (grant no. NNX15AB06G).
All computations are carried out using LANL's Institutional Computing resources.
We thank C. Dullemond for making RADMC-3D available and for useful discussions. 
Discussions with L. Perez are gratefully acknowledged as well. 
We thank the referee for comments that helped to improve the manuscript.


}


\begin{thebibliography}{}
\bibitem[ALMA Partnership et al.(2015)]{Partnership2015} Partnership, A., Brogan, C.~L., Perez, L.~M., et al.\ 2015, \apjl, 808, L3 

\bibitem[Andrews et al.(2009)]{Andrews2009} Andrews, S. M. et al. \ 2009, \apj, 700, 1502 

\bibitem[Butler et al.(2006)]{Butler2006} Butler, R.~P., Wright, J.~T., Marcy, G.~W., et al.\ 2006, \apj, 646, 505 


\bibitem[Close et al.(1997)]{Close1997} Close, L.~M., Roddier, F., Hora, J.~L., et al.\ 1997, \apj, 489, 210 

\bibitem[Crida, Morbidelli, \& Masset (2006)]{Crida2006} Crida, A., Morbidelli, A., \& 
Masset, F.\ 2006, ICARUS, 181, 587

\bibitem[Dipierro et al.(2015)]{Dipierro2015} Dipierro, G., Price, D., Laibe, G., et al.\ 2015, arXiv:1507.06719 

\bibitem[Dong et al.(2014)]{Dong2014} Dong, R., Zhu, Z., \& Whitney, B.\ 2014, arXiv:1411.6063 

\bibitem[Dullemond et al.(2002)]{Dullemond2002} Dullemond, C.~P., van Zadelhoff, G.~J., \& Natta, A.\ 2002, \aap, 389, 464 


\bibitem[Fu et al.(2014)]{Fu2014} Fu, W., Li, H, Lubow, S.,  Li, S., \& Liang, E. \ 2014, ApJL, 795, L39

\bibitem[Gonzalez et al.(2015)]{Gonzalez2015} Gonzalez, J.-F., et al., \ 2015, MNRAS, in press 
(arXiv:1509.00691)

\bibitem[Guilloteau et al.(2011)]{Guilloteau2011} Guilloteau, S., Dutrey, A., Pi{\'e}tu, V., \& Boehler, Y.\ 2011, \aap, 529, A105 

\bibitem[Isella, Carpenter \& Sargent(2009)]{Isella2009} Johansen, A., \& Klahr, H. \ 2005, ApJ, 634, 1353

\bibitem[Johansen \& Klahr(2005)]{Johansen2005} Isella, A., Carpenter, J. M., 
\& Sargent, A. I.  \ 2009, ApJ, 701, 260

\bibitem[Kataoka et al.(2015)]{Kataoka2015} Kataoka, A., Muto, T., Momose, M., Tsukagoshi, T., \& Dullemond, C.~P 2015, arXiv:1507.08902 

\bibitem[Kwon et al.(2011)]{Kwon2011} Kwon, W., Looney, L.~W., \& Mundy, L.~G.\ 2011, \apj, 741, 3 

\bibitem[Kwon et al.(2015)]{Kwon2015} Kwon, W., Looney, L.~W., Mundy, L.~G., \& Welch, W.~J.\ 2015, \apj, 808, 102

\bibitem[Li et al.(2000)]{Li2000} Li, H., Finn, J.M., Lovelace, R.V.E.,  \&  Colgate, S.A. \ 2000, 
\apj, 533, 1023

\bibitem[Li et al.(2005)]{Li2005} Li, H., et al. \ 2005, \apj, 624, 1003

\bibitem[Li et al.(2009)]{Li2009} Li, H., Lubow, S. H., Li, S., \& Lin, D. N. C. \ 2009, ApJL, 690, L52

\bibitem[Lin et al.(1994)]{L94} Lin, D.N.C., et al.  \ 1994, ApJ, 435, 821

\bibitem[Lovelace et al.(1999)]{Lovelace1999} Lovelace, R.V.E., Li, H., Colgate, S.A., \& Nelson, A.F. \ 1999, 
\apj, 513, 805

\bibitem[Mathis, Rumpl, \& Nordsieck(1977)]{mnr77} Mathis, J. S., Rumpl, W. \& Nordsieck, K. H. \ 1977, 
\apj, 217, 425

\bibitem[Pinilla et al.(2012)]{Pinilla2012} Pinilla, P., Birnstiel, T., Ricci, L., et al.\ 2012, \aap, 538, A114 

\bibitem[Pinte et al.(2015)]{Pinte2015} Pinte, C., Dent, W.~R.~F., Menard, F., et al.\ 2015, arXiv:1508.00584 

\bibitem[Pollack et al.(1994)]{p94} Pollack, J. B., et al. \ 1994, \apj, 421, 615 

\bibitem[Reg{\'a}ly et al.(2012)]{Regaly2012} Reg{\'a}ly, Z., Juh{\'a}sz, A., S{\'a}ndor, Z., \& Dullemond, C.~P.\ 2012, \mnras, 419, 1701 

\bibitem[Robitaille et al.(2007)]{Robitaille2007} Robitaille, T.~P., Whitney, B.~A., Indebetouw, R., \& Wood, K.\ 2007, \apjs, 169, 328 

\bibitem[Ruge et al.(2013)]{Ruge2013} Ruge, J.~P., Wolf, S., Uribe, A.~L., \& Klahr, H.~H.\ 2013, \aap, 549, A97 

\bibitem[Sargent \& Beckwith(1991)]{Sargent1991} Sargent, A.~I., \& Beckwith, S.~V.~W.\ 1991, \apjl, 382, L31 

\bibitem[Shen \& Turner(2008)]{Shen2008} Shen, Y., \& Turner, E.~L.\ 2008, \apj, 685, 553 

\bibitem[Tamayo et al.(2015)]{Tamayo2015} Tamayo, D., Triaud, A.~H.~M.~J., Menou, K., \& Rein, H.\ 2015, \apj, 805, 100 

\bibitem[Testi et al.(2014)]{Testi2014} Testi, L., Birnstiel, T., 
Ricci, L., et al.\ 2014, Protostars and Planets VI, 339


\bibitem[Weingartner \& Draine(2001)]{wd01} Weingartner, J.~C., \& Draine, B.~T.\ 2001, \apj, 548, 296 

\bibitem[White \& Hillenbrand(2004)]{White2004} White, R.~J., \& Hillenbrand, L.~A.\ 2004, \apj, 616, 998 

\bibitem[Zhang et al.(2015)]{Zhang2015} Zhang, K., Blake, G.~A., \& Bergin, E.~A.\ 2015, \apjl, 806, L7 

\bibitem[Zhang et al.(2014)]{Zhang2014} Zhang, X., Li, H., Li, S., \& Lin, D.N.C.\ 2014, \apjl, 789, L23 

\bibitem[Zhu et al.(2015)]{Zhu2015} Zhu, Z., Stone, J.M., \& Bai, X.\ 2015, \apj, 801, article id. 81

\bibitem[Zubko et al.(1996)]{z96} Zubko, V. G., et al. \ 1996, \mnras, 282, 1321

\end{thebibliography}
\end{document}